\documentclass{ar-1col-S2O}
\usepackage[numbers]{natbib}
\usepackage{url}
\usepackage{amsmath}
\jname{Xxxx. Xxx. Xxx. Xxx.}
\jvol{AA}
\jyear{YYYY}
\doi{10.1146/((please add article doi))}
\newcommand{\namesep}{\enspace\textbullet\enspace} 

\begin{document}

\markboth{Paliwal\namesep{Toscano}\namesep{Willitsch}}{Control of chemical reactions in radiofrequency ion traps}

\title{Control of chemical reactions in radiofrequency ion traps}

\author{Prerna Paliwal, Jutta Toscano and Stefan Willitsch
\affil{Department of Chemistry, University of Basel, Klingelbergstrasse 80, Basel 4056, Switzerland; email: stefan.willitsch@unibas.ch}}

\begin{abstract}
Over the past years, radiofrequency ion traps have become an attractive platform for studying chemical reactions as they enable a high degree of control over ion--molecule dynamics. In this review, we summarize techniques for the trapping and cooling of atomic and molecular ions in radiofrequency traps including Doppler and resolved‐sideband laser cooling, sympathetic cooling, and cryogenic buffer‐gas methods. We discuss strategies for controlling key reaction parameters: the preparation of specific internal quantum states by internal cooling, optical pumping, state‐selective photoionization and quantum‐logic spectroscopy; the manipulation of collision energies
through micromotion control, dynamic trapping and combination with molecular beams; 
and the selection of molecular structure via isotopic substitution, conformational separation and isomer‐specific ion generation. We illustrate applications of these approaches by discussing studies on quantum-state‐dependent kinetics, quantum‐resonance effects and structure‐sensitive reactivity in ion‐neutral collisions. We conclude by outlining future challenges, including full state‐to‐state reaction mapping, reaching the ultracold quantum regime free of micromotion, and the exploration of complex and chiral systems. 

\end{abstract}

\maketitle

\tableofcontents

\section{INTRODUCTION}

Over the past decades, ions in the gas phase confined in radiofrequency (RF) traps \cite{gerlich92a, ghosh95a, major05a} have emerged as a leading experimental platform for a variety of experiments in atomic, molecular and chemical physics. The ability to trap ions for extended periods of time (seconds to years), to prepare them over a broad range of temperatures (ranging from microkelvin to hundreds of Kelvin) and to localize them in space at temperatures close to the zero-point forms the basis for sophisticated schemes for their manipulation and control with high precision on the single-particle level. Consequently, trapped ions count among the best-controlled quantum systems which find a variety of applications in quantum computing \cite{bruzewicz19a}, precision measurements \cite{ludlow14a} and high-resolution mass spectrometry \cite{march05a}. 

In the context of chemistry, the advanced capabilities to manipulate trapped ions open up intriguing possibilities for the investigation of chemical processes \cite{willitsch08b,willitsch12a,willitsch17a}. Reaction rates, reaction pathways, product branching ratios and chemical dynamics often depend on the internal, i.e., electronic, vibrational, and rotational quantum states, kinetic energies and molecular geometries of the reaction partners. The precise control of these parameters paves the way for characterizing their detailed role in chemical reactions and thus ultimately for controlling chemical processes. 

The present review is devoted to the discussion of recent methodological advances for controlling the internal and external degrees of freedom of atomic and molecular ions in RF traps in the context of studying chemical reactions, with special focus on the past ten to fifteen years. For earlier work, broader scope or more specialised accounts, we refer the interested reader to related reviews \cite{willitsch17a,deiss24a,Karman24a,Softley23a,Lous22a,heazlewood21a,Heazlewood21b,Niranjan21a,Toscano20a,Snyder19a,zhang17a,bohn17a,balakrishnan16a,Thompson15a,heazlewood15a,gerlich06a,OHair06a,Gronert05a}. Beyond radiofrequency ion traps, reactions of ions are also being investigated using a variety of complementary methods including Rydberg atoms \cite{Martins25a,Zhelyazkova21a,Berngruber24a,Engel24a, allmendinger16a}, ion beams \cite{Friese25a,Richardson24a,Mancini24a,Fenn24a,ArmentaButt22a,Bowen21a,Hillenbrand22a}, ion storage rings \cite{Bogot25a,Gatchell25a,Grussie24a,Grussie24b}, electrostatic ion beam traps \cite{Sharma23a,Shahi19a} and selected-ion flow-tube (SIFT) experiments \cite{Lewis25a,OmezzineGnioua24a,Swift23a}. Recent reviews on these complementary methodologies include \cite{Smith25a,Wester22a,Liu22a,Zhelyazkova22a,Wanczek22a,Marlton22a,Meyer17a,jankunas15a,stuhl14a,smith11b,snow08a,Armentrout00a}.

\section{TRAPPING AND COOLING OF IONS}

\subsection{Trapping ions in radiofrequency traps}
\subsubsection{Paul traps} Quadrupole RF ion traps, also called Paul traps, employ a combination of static and RF electric fields to confine charged particles \cite{major05a}. The most widely used variant are linear quadrupole RF traps which typically consist of four electrodes arranged in a quadrupolar configuration (Fig. \ref{fig1} (a)). RF voltages $V_\text{RF}$ are applied to two opposing electrodes to generate dynamic confinement in the radial direction perpendicular to the longitudinal trap axis, whereas static potentials $V_{DC}$ are applied to ``end-cap" electrodes for trapping along the axial direction. Within an adiabatic approximation \cite{gerlich92a, major05a}, this electrode configuration generates an effective, nearly harmonic pseudopotential in which ions can be stably trapped and manipulated. Under these conditions, the motion of a trapped ion can be decomposed into two components, a slow ``secular'' motion corresponding to the ion trajectory within the pseudopotential and a fast oscillating ``micromotion'' driven by the RF field. Linear Paul traps offer several practical advantages. The micromotion vanishes along the central axis for symmetry reasons which is important for ground-state cooling and the precise control of the ion energies (see Secs. \ref{sec:cooling} and \ref{sec:ecoll}). Additionally, their geometry provides facile optical access for manipulation and detection of the ions with laser beams. This design also allows an efficient spatial overlap with neutral reactants for ion--neutral collision studies.

\subsubsection{Multipole traps}
\begin{figure}[h]
\includegraphics[width=1\linewidth]{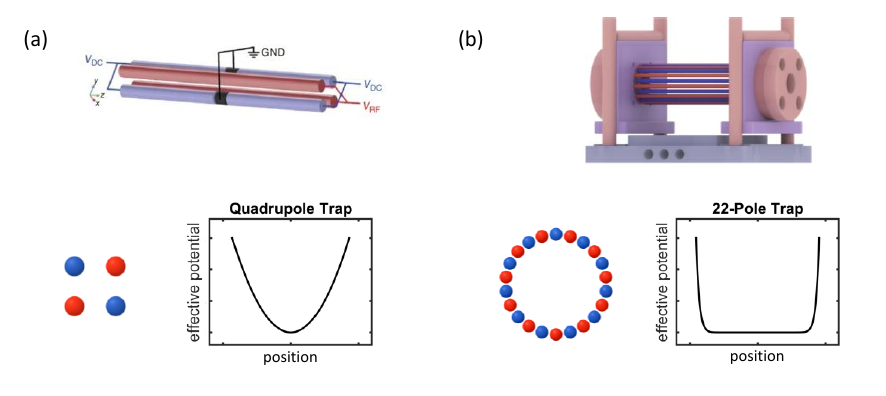}
\caption{Geometries of typical linear RF ion traps with their effective radial trapping potentials: (a) Linear quadrupole trap, (b) 22-pole trap. The colored circles depict a cross section of the electrodes configuration perpendicular to the longitudinal axis of the trap. See text for details.}
\label{fig1}
\end{figure}

Multipole RF traps extend the concept of linear quadrupole traps to a larger number of RF electrodes thus forming hexapole, octopole, or even 22-pole geometries (Fig. \ref{fig1} (b)) to shape the trapping potential \cite{gerlich92a, wester09a}. Contrary to the quadrupole field of the Paul trap, which results in a nearly harmonic confinement, multipole traps exhibit softer pseudopotentials in the radial direction with reduced micromotion across an extended region around the trap center (see Fig. \ref{fig1} (b)). However, optical access is often reduced in these traps and the non-harmonic nature of the trapping potential complicates the theoretical modeling of ion motion compared to quadrupole traps \cite{gerlich92a}. Moreover, the weaker ion confinement renders laser cooling and Coulomb crystallization inefficient (see below).

\subsection{Cooling of trapped ions}
\label{sec:cooling}

\begin{figure}[h]
\includegraphics[width=1\linewidth]{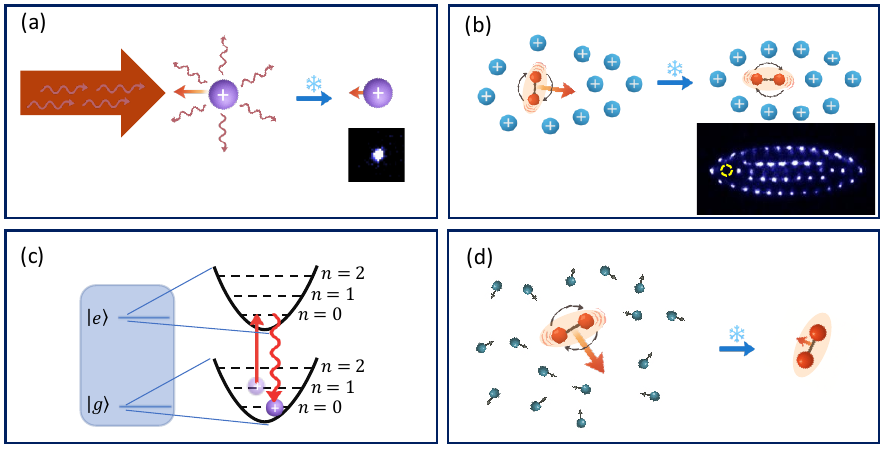}
\caption{Methods for the cooling of ions in RF traps: (a) Doppler laser cooling (inset: false-color fluorescence image of a single laser-cooled Ca$^+$ ion). (b) Sympathetic cooling of molecular ions by laser-cooled atomic ions (inset: a Coulomb crystal of laser cooled Ca$^+$ ions containing a single dark sympathetically cooled N$_2^+$ ion whose position is indicated by the yellow circle). (c) Resolved-sideband cooling on a red motional sideband of an optical transition between two levels $|g\rangle$ and $|e\rangle$ in the quantum regime of ion motion. (d) Collisional cooling with a cryogenic buffer gas.}
\label{fig2}
\end{figure}

\subsubsection{Doppler laser cooling} 
Doppler cooling is the simplest form of laser-based methods used to reduce the kinetic energy of trapped ions \cite{foot05a,metcalf99a}. Ions moving opposite to the propagation direction of a laser beam with a wavelength slightly red-detuned from an optical transition absorb photons more frequently than particles moving in the same direction as the laser beam due to the Doppler effect (see Fig. \ref{fig2} (a)). Thus, the directed momentum transfer from photons absorbed from the laser beam during repeated absorption-emission events produce a net cooling effect on the ion. This approach enables efficient cooling of ions down to temperatures of around 1 millikelvin, limited by the natural linewidth of the spectroscopic transition (the ``Doppler limit") \cite{metcalf99a}. At these temperatures and under sufficiently strong confinement as can be afforded by quadrupole traps, the ions undergo Coulomb crystallization \cite{drewsen02a, willitsch08a, willitsch12a}, i.e., they localize in the trap to form ordered structures (see insets in Fig. \ref{fig2} (a) and (b)). Under these conditions, it is possible to observe, address and manipulate single particles. 

\subsubsection{Sympathetic cooling}
Doppler laser cooling requires the repeated scattering of photons from the ion. Therefore, the applicability of Doppler cooling has thus far been restricted to atomic ions with sufficiently simple energy-level structures (e.g., Be$^+$, Mg$^+$, Ca$^+$, Ba$^+$, Yb$^+$, Sr$^+$) which exhibit closed optical cycling transitions. Up to now, the direct laser cooling of a molecular ion has not been achieved (in contrast to a range of neutral molecules \cite{tarbutt18a}).
In order to cool trapped molecular ions which do not feature closed-cycling transitions, sympathetic cooling has emerged as a widely used technique. In this approach, molecular ions are trapped together with laser-cooled atomic ions, and kinetic energy is transferred from the molecular to the atomic species through Coulomb interactions (Fig. \ref{fig2} (b)). As a result, the molecular ions reach translational temperatures close to their atomic collision partners and Coulomb-crystallize in the trap. While this is a highly efficient method to cool the translational motion, the isotropic, long-range nature of the Coulomb interaction between the ions prevents coupling to the internal molecular degrees of freedom \cite{bertelsen06a} unless the collision energies become very high \cite{berglund24a}. Therefore, sympathetic cooling has to be combined with alternative methods to control the internal degrees of freedom, as discussed in Sec. \ref{sec:stateprep}. 

\subsubsection{Resolved-sideband cooling} In the quantum regime, the motion of an ion trapped in the harmonic pseudopotential of a quadrupole RF trap can be modeled as a quantum harmonic oscillator. At very low temperatures reached after Doppler cooling and under strong confinement where the spatial extent of the ion motional wavefunction is much smaller than the wavelength of the cooling laser (the ``Lamb--Dicke regime") \cite{wineland98a,leibfried03a}, motional sidebands on narrow optical transitions of the ion can be spectroscopically resolved. These sidebands entail a change of the motional quantum state alongside the internal state of the ion. 
A laser beam tuned to a ``red sideband" of an optical transition thus simultaneously excites the internal state while decreasing the motional quantum number of the ion (Fig. \ref{fig2} (c)). Repeated application of this process, combined with spontaneous emission that preserves the motional state, results in the systematic removal of quantized motional energy from the ion and its cooling to the ground state of motion in the trap. The associated zero-point energy represents the physical limit for cooling a trapped ion. 

\subsubsection{Cryogenic buffer gas cooling}
Collisional cooling with a cryogenic buffer gas is a widely used method for reducing both the kinetic and internal energy of trapped ions, particularly in cases where laser cooling is not feasible. In this technique, a cryogenic neutral gas, commonly helium, is introduced into the ion trap. The ions undergo collisions with the buffer-gas atoms, leading to thermalization close to the gas temperature, typically in the range around ten Kelvin \cite{wester09a} (Fig. \ref{fig2} (d)). Buffer-gas cooling is compatible with a broad range of atomic and molecular species and is typically employed in higher-order multipole RF traps in which RF heating caused by collision-induced coupling of micromotion and secular motion is suppressed owing to the reduced micromotion close to the trap center \cite{hoeltkemeier16a} (see also Sec. \ref{sec:ecoll}). However, it generally results in higher translational temperatures compared to laser-based cooling techniques and is not state selective as it leaves rotational and vibrational populations thermalized.

\section{CONTROLLING REACTION PARAMETERS IN TRAPPED-ION EXPERIMENTS}
        
\subsection{Quantum states}
  \label{sec:stateprep}
            
One of the key properties influencing the kinetics, dynamics and outcome of a chemical reaction is the quantum state of the reactants. Consequently, the control of the quantum state of the collision partners has been one of the longstanding aims, if not a holy grail, of chemical reaction dynamics \cite{pan23a, willitsch17a}. While efforts in the context of trapped-ion experiments have mainly concentrated on atomic species with comparatively simple energy-level structures in the past, experimental methodology has now progressed to a point at which more complex quantum systems such as molecules have moved into focus and can now be controlled with a precision approaching the one achieved in atoms \cite{sinhal23a}.

\subsubsection{Reactions with state-selected atoms and atomic ions in traps}
 
Within the realm of ion--neutral collision studies, hybrid experiments which combine trapped ions with trapped ultracold atoms \cite{deiss24a, tomza19a, sias14a, haerter14a, willitsch15a} have so far reached the most sophisticated level of control. In these experiments, laser-cooled atomic ions in RF traps are superimposed with ultracold atoms prepared in magnetic, magneto-optical (MOT) or optical-dipole traps. Using optical pumping methods, both the atomic ions and neutral atoms can be prepared in specific quantum states to study state-controlled collisions at temperatures down to the mK, and recently even $\mu$K, regimes.

Early studies focused on electronic-state-specific ion--atom reaction studies by adjusting the parameters for the laser cooling, and therefore the electronic-state populations, of both the ions and atoms \cite{hall11a, hall13b, rellergert11a, sullivan12a, haze15a, krukow16b, joger17a}. These investigations concentrated on electronic-state-specific kinetics, in particular the speed-up of charge-exchange reactions upon electronic excitation, and the formation of molecular ions by radiative association. Subsequently, the engineering of excited-state interactions and, therefore, of the reaction kinetics was demonstrated by dressing the collision system with laser fields \cite{mills19a}. Beyond electronic states, in some experiments both the atomic ions and neutral atoms were prepared in specific hyperfine and even Zeeman states \cite{ratschbacher12a, sikorsky18a, feldker20a} to explore state-specific spin-relaxation and spin-exchange processes. Such studies revealed signatures of collisional quantum effects such as evidence for scattering resonances \cite{feldker20a, weckesser21a, thielemann25a} as well as the phase-locking of collisional wavefunctions even at comparatively high energies \cite{sikorsky18b, walewski25a}. Transgressing beyond atomic ions, reactions between sympathetically cooled molecular ions and ultracold atoms have elucidated the role of the electronic state and the importance of higher-order long-range intermolecular interactions in cold molecular collisions \cite{hall12a, doerfler19a, puri17a, mohammadi21a}.

While ion--atom hybrid-trapping experiments predominantly operate in the mK regime, the state control achieved in laser-cooled, Coulomb-crystallized atomic ions has also been harnessed in studies of the state-dependence of reactions with ``warmer" molecular gases. For instance, the dependence of reaction rates of halomethanes with state-controlled laser-cooled Ca$^+$ ions was characterized in Refs. \cite{willitsch08a, gingell10a}. These studies explored the role of submerged energy barriers in ion--molecule reactions which were shown to suppress reactions with ground-state ions while reaction rates were found to accelerate upon electronic excitation. A similar scenario was uncovered in reactions of laser-cooled Be$^+$ with H$_2$O \cite{yang18a}. Laser cooling has also been used to study the state dependence of rate constants of reactions of Ca$^+$ with molecular oxygen \cite{drewsen02a, schmid19a} revealing that the reaction only occurs with electronically excited ions. Similarly, the reactions of Ca$^+$ \cite{kimura11a}, Mg$^+$ \cite{molhave00a} and Be$^+$ \cite{roth06b} with H$_2$ as well as Ca$^+$ \cite{okada03a, wu24a} with H$_2$O were shown to proceed only out of electronically excited states. 

\subsubsection{Reactions with state-selected molecules and molecular ions in traps}

\begin{figure}[h]
\includegraphics[width=1\textwidth]{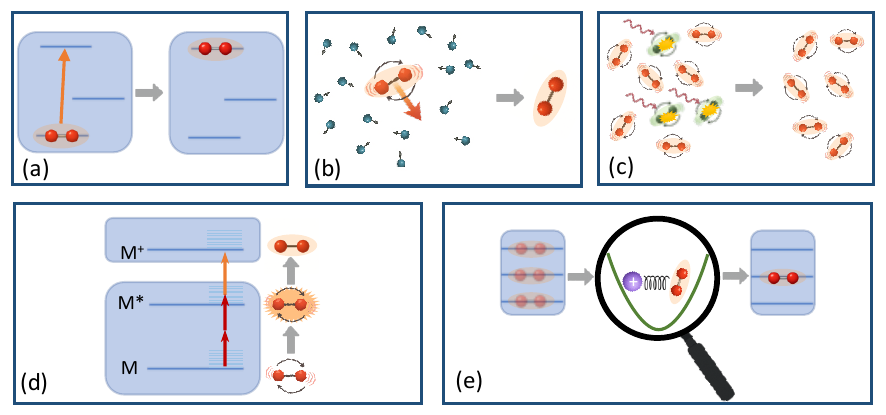}
\caption{Methods for the internal-state preparation of molecular ions: (a) Optical pumping into selected states. (b) Buffer-gas cooling of the internal degrees of freedom. (c) State-selective destruction of molecular ions by removing specific quantum states from the trap. (d) Resonance-enhanced multiphoton threshold photoionization for generating molecular ions in specific rovibrational states. (e) Projective state preparation of single molecular ions through state measurement (symbolized by the magnifying glass) using quantum-logic spectroscopy.}
\label{figstateprep}
\end{figure}

Unlike atomic ions, molecular ions possess a rich internal structure arising from vibrational and rotational degrees of freedom. Thus, in addition to the spin and electronic levels, the rovibrational state can influence reactivity. 

\paragraph*{Vibrational effects:}Vibrational effects can be probed by the preparation of molecular ions in specific vibrational states using optical pumping (Fig. \ref{figstateprep} (a)). This approach has, for example, been pursued in recent studies using a cryogenic 22-pole RF ion trap with infrared laser excitation of C$_3$H$_2^+$ ions \cite{markus20a}. In that study, a buffer gas of molecular hydrogen was continuously introduced into the trap, providing thermalization through collisions and also serving as the neutral reactant. A narrow-band IR laser populated specific rovibrational levels in the ion and the proton transfer reaction between C$_3$H$_2^+$ and H$_2$ was then monitored by mass spectrometry. Surprisingly, excitation of the $\nu_7$ antisymmetric C–H stretch vibration significantly reduced the reaction rate, which was attributed to the inhibited formation of a weakly bound complex in the entrance channel of the reaction.

Similar optical-pumping approaches have been employed for the preparation of vibrational states in studies of the vibrational quenching of molecular anions. In a recent study, C$_2^-$ anions were cooled in a cryogenic ion trap using helium buffer gas and optically pumped into the $v = 1$ vibrational level \cite{notzold23a}, followed by measurements of the absolute quenching rate via collisions with H$_2$. The observed rate for the $v = 1 \rightarrow v = 0$ transition was found to be more than three orders of magnitude smaller than the classical Langevin capture rate.

\paragraph*{Rotational effects:}A variety of approaches have been implemented for controlling rotational-state populations in molecular ions and thus studying rotational-state effects. A simple method relies on tuning rotational-state populations by changing the internal temperature of the ions (Fig. \ref{figstateprep} (b)). Such an approach was pursued, for instance, to probe the effect of rotational excitation in the CH$^+ +$ H $\rightarrow$ C$^+ +$ H$_2$ reaction \cite{plasil11b} in a cryogenic 22-pole trap. 
The CH$^+$ ions were rotationally cooled by collisions with helium buffer gas at a variable temperature inside the trap, allowing the preparation of different distributions of rotational-level populations and, therefore, the exploration of rotational-state-dependent effects. 
It was found that the reactivity was strongly suppressed for CH$^+$ in its rotational ground state which still needs to be rationalized by theoretical models \cite{faure17a}.

In a complementary approach to rotational-state-resolved studies, the reactivity of trapped SiO$^+$ ions was explored in linear Paul trap using optical-pumping methods for internal-state control \cite{venkataramanababu23a}. SiO$^+$ ions 
were prepared in narrowly defined intervals of rotational states using broadband optical pumping allowing access to both low-$J$ levels and super-rotor states with rotational quantum numbers up to $J = 170$. The ion translational motion was sympathetically cooled using laser-cooled Ba$^+$ ions. 
The hydrogen abstraction reaction SiO$^+ +$ H$_2 \rightarrow$ SiOH$^+$ + H was then studied by leaking H$_2$ gas into the vacuum chamber. 
A three-fold increase in reactivity observed for highly excited rotational states was attributed to strong coupling between SiO$^+$ rotational motion and the reaction coordinate at the transition state consistent with predictions from quasi-classical trajectory simulations.

Optical pumping can also be used for cooling the rotational motion.``Rotational laser cooling" using mid-infrared laser sources has been performed by driving selected rovibrational transitions that subsequently lead, via spontaneous emission and population redistribution by ambient blackbody radiation, to an accumulation of population in the rovibrational ground state. For example, experiments with trapped MgH$^+$ ions achieved a 15-fold increase in rotational ground-state population reaching 36$\%$ as in a 20 K thermal distribution \cite{staanum10a}. A similar scheme was applied to HD$^+$ ions using two continuous-wave lasers which optically pumped the population into the rovibrational ground state, achieving 78$\%$ efficiency, corresponding to a rotational temperature of 26.5 K within 40 s \cite{schneider10a}. A different approach was demonstrated for AlH$^+$ ions using broadband rotational optical cooling. By driving multiple rotational transitions simultaneously with spectrally filtered femtosecond laser pulses in resonance with an electronic transition, AlH$^+$ ions were cooled from room temperature to the rotational ground state increasing the ground-state population to over 95$\%$ \cite{lien14a}. 

Another approach to state preparation relies on state-selective destruction where specific internal states are removed rather than populated (Fig. \ref{figstateprep} (c)). This method was applied in studies of rotationally inelastic collisions between hydroxide anions (OH$^-$) and helium atoms. OH$^-$ ions were confined and thermalized in a cryogenic multipole trap yielding a well-characterized initial rotational distribution \cite{hauser15a}. To eliminate ions in the $J = 1$ rotational level, a photodetachment laser was tuned to selectively remove OH$^-$($J = 1$), leaving only $J = 0$ ions for subsequent collision experiments. A related method is ``leak-out spectroscopy" in which excitation of specific rovibrational transitions and subsequent vibrational-to-translational energy transfer in collisions with buffer gas causes ions to gain sufficient kinetic energy to escape a shallow trap. This scheme acts both as a state filter and a highly sensitive spectroscopic probe of state populations \cite{schmid22a}.

Expanding the toolkit for state-resolved reaction studies, combining a Coulomb-crystal environment with resonance-enhanced multiphoton threshold photoionization (REMPI) for ion preparation (Fig. \ref{figstateprep} (d)) \cite{tong10a, shlykov23a} and laser-induced charge transfer (LICT) \cite{schlemmer99a} for state-selective detection have enabled detailed investigations of rotational ion–molecule reaction dynamics. In studies of the charge-transfer reaction N$_2^+$ + N$_2$ → N$_2$ + N$_2^+$, N$_2^+$ ions were generated in rovibrational ground state using REMPI and confined in a linear Paul trap within laser-cooled Ca$^+$ Coulomb crystals, providing sympathetic cooling \cite{tong12a}. Neutral N$_2$ molecules from a supersonic beam were introduced to initiate collisions. The rotational state distribution of the product N$_2^+$ ions was probed using LICT. This approach allowed for direct observation of translation-to-rotation energy transfer and the elucidation of relative state-to-state rate constants.

Projective state preparation (Fig. \ref{figstateprep} (e)) using quantum-logic spectroscopy \cite{schmidt05a} represents one of the most refined techniques for controlling the state of single trapped molecular ions \cite{shlykov23a}. This method employs a co-trapped atomic ion as a quantum sensor \cite{wolf16a, chou17a, sinhal20a}, allowing for indirect yet high-fidelity readout of the molecular ion’s internal state \cite{sinhal20a, najafian20b}. For instance, in the approach of Ref. \cite{sinhal20a} an N$_2^+$ molecular ion was first prepared in the trap using REMPI. A three-step quantum logic protocol was then used to interrogate this state. First, the shared motion of a Ca$^+$--N$_2^+$ two-ion crystal was cooled to its motional ground state in the trap by resolved-sideband cooling on the atomic ion. Second, a state-dependent optical dipole force produced by focused laser beams was applied to the molecular ion inducing motion of the ions only if the molecular ion was in the targeted quantum state. Third, the excited motion was then detected on the atomic ion, whose fluorescence encoded information about the shared motion with the molecule and, therefore, the molecule’s quantum state. By measuring the state of the single molecular ion, the ion was projected (and therefore prepared) in the specific state through the measurement process \cite{shlykov23a}. This approach has been applied in the projective preparation of specific rovibronic states of N$_2^+$ which were subsequently subjected to inelastic and reactive collisions \cite{najafian20b}.

An important step in controlling reaction dynamics involves extending quantum-state selectivity to the neutral collision partner. In this context, recent advances in molecular-beam manipulation techniques have made it possible to control the rotational states of neutral molecules. For example, electrostatic deflection was used to separate different rotational states of polar neutral molecules based on differences in the Stark effect \cite{chang15a}. For the case of water molecules, this method enabled the separation of its two nuclear-spin isomers, \emph{para} and \emph{ortho}, which are associated with distinct rotational states \cite{horke14a}. Capitalizing on this methodology, beams of electrostatically deflected \emph{ortho}- and \emph{para}-water were recently directed towards Coulomb-crystallized diazenylium ions (N$_2$H$^+$) confined in a linear quadrupole ion trap \cite{kilaj18a}. The experiments demonstrated a dependence of chemical reactivity on rotational state where \emph{para-}water, in the rotational ground state, reacted faster than \emph{ortho-}water in the first excited rotational state. This difference in reactivity was attributed to varying degrees of rotational averaging in long-range ion--dipole interactions.

In another study, a cryogenic helium-buffer-gas cell was combined with a Stark velocity filter to produce a rotationally cooled, velocity-selected beam of polar CH$_3$F molecules \cite{okada22a}. This beam was subsequently introduced into a cryogenic linear Paul trap containing laser-cooled Ca$^+$ ions. The experiment revealed a dependence of the reaction rate on the rotational population distribution of CH$_3$F, with molecules exhibiting colder rotational distributions showing enhanced reactivity. These findings are rationalized in terms of Perturbed Rotational State (PRS) theory. 

\subsection{Collision energy}

\label{sec:ecoll}

\begin{figure}[h]
\includegraphics[width=1\linewidth]{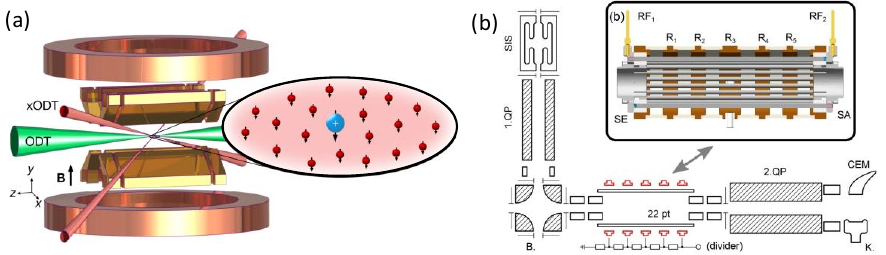}
\caption{Representative experimental setups for studying collisions and reactions of neutral species with trapped ions: (a) Hybrid trapping experiment combining a ``blade" linear-quadrupole RF trap with an optical-dipole trap (ODT) for confining ultracold atoms \cite{weckesser21a}. (b) Cryogenic 22-pole trap setup from Ref. \cite{Jusko24a} consisting of a storage ion source (SIS) as well as quadrupole guides (QP) and benders (B) connecting the trap (22pt) with the source and particle detector (CEM). Panel (a) is reproduced from Reference \cite{weckesser21a} with permission from Springer Nature and P. Weckesser. Panel (b) is reproduced from Reference \cite{Jusko24a} (CC BY 4.0).} 
\label{fig4}
\end{figure}

Apart from the quantum state, the collision energy of the reaction partners is one of the key parameters influencing the dynamics of chemical processes. The collision-energy dependence of a reaction encodes important information on, for example, intermolecular interactions, quantum effects such as resonances, reaction barriers and mechanisms. At the same time, this is one of the most difficult parameters to control in a trapped-ion setting (Fig. \ref{fig4}). In an ideal experiment, collision energies can be varied continuously over a wide range while maintaining a narrow energy spread of the particles, as has recently been realized for neutral species using Stark- and Zeeman-decelerated or merged molecular beams \cite{henson12a, jankunas15a, tang23a, allmendinger16a}. However, a trap necessarily entails the immobilization of the trapped particles. In Coulomb crystals, the ions are strongly localized, and if at all, diffuse slowly through the crystal. The kinetic energy of the ions is then usually dominated by their micromotion which depends on their position in the trap \cite{major05a}. Because the micromotion is a driven oscillatory motion, the total energy of the ions is time-dependent. Ions in RF traps are therefore not in thermal equilibrium and care must therefore be exercised when assigning temperatures to the trapped particles and their collisions. While unperturbed secular motion is in good approximation thermal and can be described by a Maxwell-Boltzmann distribution, collisions of the ions with buffer or background gas particles couple micro- and secular motion. In this case, the ion secular energies obey Tsallis or related distributions instead \cite{meir16a, rouse17a, rouse18a, rouse19a}.

On the central symmetry axis of a linear trap, the RF fields vanish for symmetry reasons. Therefore, ions located on the trap axis have zero micromotion energy. In bigger crystals, however, a number of ions are invariably located off this RF null line. 
This situation puts stringent constraints on the energies of the ions. Big crystals usually exhibit a broad energy distribution ranging from almost zero to up to hundreds of Kelvin \cite{bell09a, hall13a}. By changing crystal size and shape or by shifting crystals away from the RF null line, the micromotion imparted to the ions can be modified and the energy distributions can be tuned. This approach has been used in several studies to vary collision energies in interactions with stationary clouds of trapped ultracold atoms in order to determine collision-energy-dependent reaction cross sections \cite{hall13a, hall12a, xiao24a}. In this way, for instance, the effect of charge--quadrupole interactions in ion--atom collisions could be deduced from the energy dependence of the reaction rates at collision energies in the range of tens of millikelvin \cite{hall12a}.

The lowest energies can be reached in experiments with single ions which are placed on the RF null line. In this case, ion energies are limited by the precision of the minimization of excess micromotion caused by stray electric fields which push the ion away from the trap centre \cite{berkeland98a}. Thus far, collision energies in the regime of tens of microkelvins have recently been demonstrated in experiments with Li/Yb$^+$ mixtures \cite{feldker20a} and in Li/Ba$^+$ mixtures \cite{weckesser21a, thielemann25a}. At such exceptionally low energies, quantum signatures, i.e., scattering resonances, could be discovered for the first time in ion--atom collisions.

However, even in experiments in which single ions are dislocated from the RF null line in a precisely controlled fashion \cite{hall12a, feldker20a, thielemann25a}, their collision-energy distributions tend to be broad because of the oscillatory nature of the imparted micromotion. This is generally undesirable for precise collision experiments and makes it challenging to, for example, unravel collisional quantum effects which usually appear within narrowly defined collision-energy intervals \cite{feldker20a, weckesser21a}. Moreover, even if an ion is exactly placed on the central trap axis, interparticle interactions invariably dislodge the particle from the RF null line during a collision \cite{cetina12a}. This effect can be minimized by favourable collision kinematics in systems with heavy ions and light neutrals, as demonstrated in the aforementioned studies using Li/Yb$^+$ \cite{feldker20a} and Li/Ba$^+$ \cite{weckesser21a}. However, these effects cannot be eliminated completely and in the end impose stringent lower bounds on the collision energy achievable in RF traps \cite{cetina12a, krych15a}.

In the context of ion--atom hybrid trapping experiments, two different approaches have been employed to realize tunable, but still well-defined and narrow, collision energy distributions: first, to keep the ions immobilized close to the RF null line and impart a well-defined kinetic energy to the atoms \cite{eberle16a, doerfler19a, voute23a, katz22a, pinkas23a, walewski25a}, and second, the other way round \cite{puri18a, puri19a}. 

The first approach has been implemented in experiments employing a ``dynamic" hybrid trap \cite{eberle16a} in which clouds of cold Rb atoms were trapped in a MOT displaced from ion crystals in the center of the trap. The cold atoms were then accelerated to a specified, but variable velocity using radiation pressure forces of a precisely defined strength exerted by one of the MOT laser beams. After passing through the ions and inducing collisions, the atom cloud was recaptured in another MOT on the other side of the ion crystal to be accelerated back to its original position, passing the ion crystal again. In this way, a cloud of cold atoms was turned into a slow, high-density atomic beam with a well-defined velocity that allowed to study cold ion--atom collisions at a specific, but variable collision energy. Using this technique, the dynamics of collisions between Rb atoms and Ca$^+$, N$_2^+$, O$_2^+$ and N$_2$H$^+$ ions was studied with collision energies from $\approx$10-50~mK with relative energy spreads of order 10--20\% \cite{eberle16a, doerfler19a, voute23a}.

An alternative implementation of this approach relied an ``optical conveyor belt" formed by a far off-resonant optical dipole trap into which ultracold Rb atoms were loaded \cite{katz22a, pinkas23a, walewski25a}. The atoms were then shuttled with a defined velocity across Sr$^+$ ions placed on the RF null line of an RF trap. Thus, hyperfine-changing collisions, the formation of trap-assisted bound ion--atom states and quantum interference effects could be studied in detail.

The second approach was demonstrated in Ref. \cite{puri18a}. In this case, a MOT for Ca atoms was kept stationary and ions were shuttled back and forth through the atom cloud using carefully tailored voltage waveforms applied to the trap's endcap electrodes. In this way, ion energies ranging from the mK to the K range could be achieved with a relative energy spread of order 10\%. This method was used to study the collision-energy dependence of reactions of cold Ca atoms with Yb$^+$, BaCl$^+$ and BaOCH$_3^+$ ions \cite{puri18a, puri17a, mills20a}. 

While hybrid-trapping experiments primarily focus on the sub-Kelvin collision-energy range, a number of alternative approaches have been implemented that enable reaction studies in traps at tunable energies in the interval from a few Kelvins up to room temperature, which is of particular interest in the context of astrochemistry \cite{herbst01b}. These experiments all rely on the combination of an ion trap, in which cold ions are prepared by laser, sympathetic or buffer-gas cooling, with a molecular-beam source delivering the neutral co-reagent with a tunable velocity.

A first class of experiments combined laser- or sympathetically cooled ions in quadrupole traps with bent electrostatic quadrupole guides \cite{willitsch08a, bell09a, okada20a, okada22a}. The bent quadrupoles act as tunable velocity filters for effusive beams of polar molecules \cite{rangwala03a}, which were then directed at Coulomb crystals of the ions in the trap. In this way, the energy dependence of reactions of atomic ions like Ca$^+$ with various polar gases such as CH$_3$F, CH$_3$Cl, CH$_2$F$_2$ \cite{gingell10a, okada22a} and CH$_3$CN and ND$_3$ \cite{okada13a} as well as of sympathetically cooled N$_2$H$^+$ molecular ions with slow CH$_3$CN an ND$_3$ \cite{okada13a} were studied. Recent evolutions of this approach involved combining the velocity selector with a cryogenic-buffer-gas beam source, which also enabled a variation of the rotational temperature of the neutral molecules \cite{okada20a, okada22a}. In this way, the effects of both rotational and translational energy in ion--molecule reactions could be independently explored.

A second class of experiments combined trapped ions with cryogenic-buffer-gas beams. Gerlich and co-workers pioneered the development of a cryogenic beam of H, H$_2$ and their isotopomers which was combined with a cryogenic 22-pole trap in which molecular ions were buffer-gas cooled \cite{borodi09a}. This approach has been used for studies of a variety of astrochemically relevant ion--molecule reactions of cold hydrogen atoms and molecules with ions such as CO$_2^+$\cite{borodi09a}, CH$^+$ \cite{plasil11b}, H$^-$ \cite{gerlich12a}, and D$^-$ \cite{rouvcka15a}. In a similar approach, a cryogenic buffer-gas beam of H$_2$O has been combined with sympathetically cooled ions in a quadrupole trap yielding insights into the low-temperature kinetics and dynamics of the reactions of water with C$^+$ ions \cite{yang21a} yielding the product isomers HCO$^+$ and HOC$^+$.
More recently, the kinetic energies of buffer-gas cooled ions confined in a 22-pole radiofrequency trap have been characterized using an “evaporation” technique. Small potential barriers generated by ring electrodes in the trap were systematically varied allowing the energy-dependent extraction of the ions \cite{Redondo25a}. This technique has been used to sample the energy of buffer-gas thermalized He$^+$ ions as well as to identify energetic H$_3^+$ ions produced in an exothermic reaction between H$_2^+$ and H$_2$ inside the trap.

Table \ref{tab1} summarises the key methods discussed in this section to control the collision energy in trapped-ion experiments.

\begin{table}[h]
\tabcolsep7.5pt
\caption{Overview of key methods for controlling ion and collision energies in ion--neutral collision experiments in traps and their typical energy ranges (in units of Kelvin).}
\label{tab1}
\begin{center}
\begin{tabular}{@{}l|c@{} | c@{}}
\hline
\textbf{Method of controlling collision energy }& \textbf{Typical energies } & \textbf{Refs.} \\
\hline

Ground-state cooling of single ions placed near RF null line & 1--100 µK& \cite{feldker20a, weckesser21a} \\

Single ions deterministically displaced from RF null line & mK--K & \cite{hall12a, xiao24a} \\

Dynamic hybrid trap (moving atom cloud with static ions) & 10--100 mK & \cite{eberle16a} \\

Optical conveyor belt (moving atom cloud with static ions) & 1--100 mK & \cite{katz22a, walewski25a}\\

Ion shuttling through static atom cloud & mK--K & \cite{puri18a} \\

Trapped ions with molecular beams or bent quadrupole guides & 1--100 K & \cite{willitsch08a, chang13a, okada20a}\\

Trapped ions with cryogenic buffer-gas beam& 1--100 K & \cite{borodi09a, yang21a}\\
\hline
\end{tabular}
\end{center}
\end{table}

\subsection{Molecular structure}


The idea that chemical reactivity depends on the specific combination of reacting species goes back to the alchemical roots of chemistry, long predating concepts such as the existence of molecules composed of atoms. Much of the diversity displayed by chemical reactions is indeed a result of the distinct {\it composition} of different molecules, consisting of different atoms which arrange themselves to form numerous different functional groups (such as alcohols, amines, alkanes etc.) which, in turn, lead to distinct properties and chemical behaviour. A more subtle but equally significant quest has been to understand how chemical reactivity is affected by the {\it structure} of molecules and, therefore, by the relative arrangement of the atoms and functional groups within them. 

\begin{figure}[h]
\includegraphics[width=12cm]{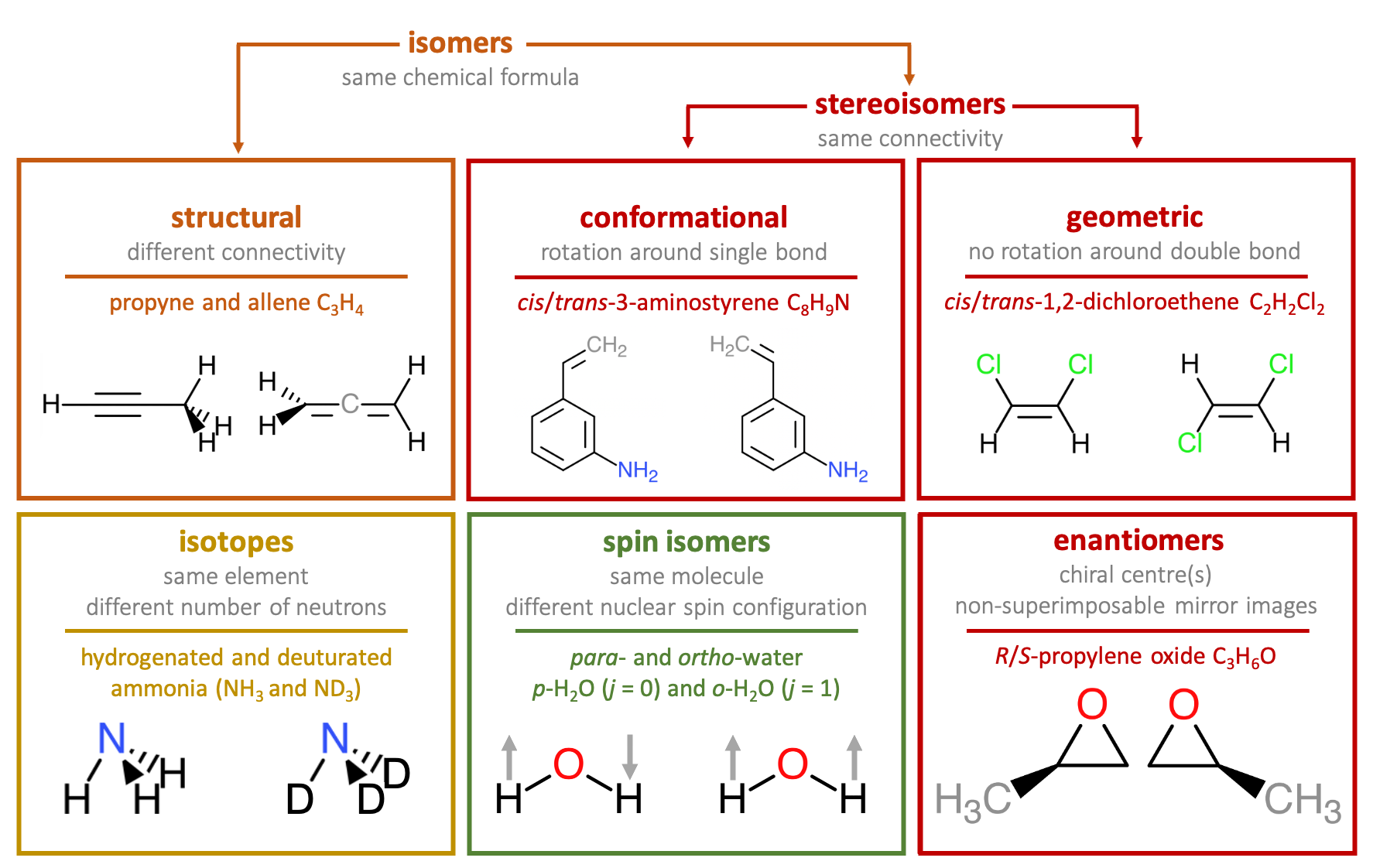}
\caption{Schematic summary of different types of isomers with examples. See text for details. 
}
\label{figMS1}
\end{figure}

Examining the influence of molecular structure independently of molecular composition is best accomplished by comparing the reactivity of molecules with the same composition but different structures, i.e., of {\it isomers} (Fig. \ref{figMS1}). 
The different connectivity in structural isomers and the different relative spatial arrangement of atoms in stereoisomers can lead to markedly distinct reactivity and reaction outcomes. The interplay between long-range electrostatic interactions and short-range effects, including complex formation, steric hindrance and dynamical bottlenecks, determine the fate of different isomeric reactants, as well as the formation of different isomeric products. In the context of trap experiments, the ability to store ionic products allows for their internal energy redistribution over the duration of trapping, radiative cooling and isomerization before detection, as well as enabling the study of secondary reactions and chain-reaction mechanisms.

Aside from isomers, molecular structure can also be changed by isotopic substitution, where one or more atoms in a molecule are exchanged for an isotope, leading to different isotopologues (Fig. \ref{figMS1}). Isotopic substitution influences vibrational frequencies but does not change the electronic structure of a molecule. Lighter isotopologues may show a propensity to react faster than their heavier counterparts due to their higher vibrational frequencies and zero-point energies ({\it kinetic isotope effect}) \cite{atkins94a}. The inverse kinetic isotope effect indicates a faster reactivity of the heavier isotopologues instead. Tunnelling probabilities depend on mass and can augment kinetic isotope effects, especially at low temperatures and when hydrogen atoms are involved. 

The recent progress in the control over collision energies and internal quantum states of the reacting species outlined in the previous sections has been accompanied by advancements in the control over molecular structure in collision experiments, as will be illustrated with a few pertinent examples from the recent literature.

\subsubsection{Isotopologues}

The molecular hydrogen isotopologues (H$_2$, HD and D$_2$) feature heavily in isotopic studies, both for their particular prominence in interstellar chemistry and for their uniquely large relative mass difference (with HD and D$_2$ being 50\% and 100\% times heavier than H$_2$, respectively), which makes them particularly sensitive probes for isotopic effects. Over the last years, several studies have investigated and compared the reactivity of these isotopologues in RF ion traps. 

A series of such measurements, carried out in cryogenic 22-pole traps, examined the reactivity of molecular-hydrogen isotopologues towards various atomic and molecular cations and anions of astrophysical relevance in the temperature range 15--300~K. 
Oxygen and carbon monoxide cations (O$^+$ and CO$^+$, respectively) were found to react slightly faster with H$_2$ than D$_2$ with temperature-independent rate coefficients close to the respective Langevin capture-rate predictions \cite{Kovalenko21a,Jusko24a}. A similar temperature-independent behavior was also observed in the case of the Cl$^+$ and HCl$^+$ cations, but with rates significantly slower than Langevin, particularly for Cl$^+$ \cite{JimenezRedondo25a}.
Nitrogen cations (N$^+$), on the other hand, displayed rate coefficients which increase with temperature for all three isotopologues, on account of the endothermicity of the reactions \cite{Plasil22a}. 

Oxygen anions (O$^-$) react with H$_2$ and D$_2$ mainly {\it via} associative detachment (to form H$_2$O/D$_2$O~+~e$^-$) with atom transfer (to form OH$^-$/OD$^-$~+~H/D) being a minor channel with a rate of a few \% of the total \cite{Plasil17a}. The associative detachment rate coefficients were found to be similar for the two reactions while the atom transfer channel shows a pronounced isotopic effect with OH$^-$ being twice as likely to form compared to OD$^-$. Such reactive behaviour was attributed to zero-point energy effects in conjunction with a barrier at short range in the linear geometry of the colliding particles. The isotopologues of ammonia NH$_3$ and ND$_3$ (Fig. \ref{figMS1}) showed a pronounced inverse kinetic isotope effect in charge-exchange reactions with the rare gas ions Xe$^+$, Kr$^+$ and Ar$^+$ sympathetically cooled within a Ca$^+$ ions Coulomb crystal inside a linear quadrupole trap \cite{Petralia20a,Tsikritea21a}. The higher reactivity of the heavier deuterated ammonia molecules was attributed to the lower vibrational frequencies, and correspondingly higher density of states, in the reaction complex facilitating intramolecular vibrational redistribution. Discrepancies with SIFT measurements of the same system \cite{Ard22a} were attributed to different rotational and translational temperature distributions probed in the two experiments \cite{Petralia22a,Hahn24a}.

The isotopologues of water H$_2$O and D$_2$O, in contrast, were found to undergo charge exchange with Xe$^+$ ions with capture-limited rates. No significant difference upon isotopic substitution was found, despite the similarities in capture-relevant properties between water and ammonia \cite{Tsikritea22a}. This behaviour was explained by the presence of a crossing between the reactants and products potential energy surfaces in the case of water, and its absence in the case of ammonia, enabling efficient capture-limited and isotope-independent charge exchange. 
No significant isotope effect was observed in the reaction of H$_2$O, HDO and D$_2$O with laser-cooled Be$^+$ ions, with all isotopologues reacting with similar rates, suggesting a significant non-statistical contribution to the reactivity which deviates from the capture limit \cite{yang18a,Chen19a}. However, the reaction with HDO revealed a propensity towards the formation of BeOD$^+$ over BeOH$^+$. This finding was rationalized in the context of the interplay between two opposing isotope-dependent factors: a faster rise in centrifugal barrier for the BeOD$^+$~+~H formation due to its lower reduced mass, and a larger availability of open channels due to its reduced vibrational frequency, with the latter factor dominating within the statistical model adopted. A subsequent experiment found BeOH$^+$ to be the only product of the reaction between sympathetically cooled BeD$^+$ and H$_2$O, with no evidence for the formation of BeOD$^+$ despite the energetic similarities \cite{yang21b}. {\it Ab initio} calculations revealed that the double atomic displacement necessary to form BeOH$^+$ is accessible through a submerged barrier, while the BeOD$^+$ formation pathway via triple displacement or oxygen insertion is dynamically unfavourable. 

The isotopic-substitution studies mentioned so far involve the isotopologues of the neutral molecules which react with the trapped ionic species. Recently, isotopic substitution of trapped molecular ions has also been explored. Various isotopologues of the diatomic H$_2^+$ ion have been shown to follow Langevin reactivity, whilst those of the triatomic H$_3^+$ ion show temperature-dependent rate coefficients, regardless of whether they are endothermic or exothermic \cite{JimenezRedondo24a}.
The isotopologues of the CCl$^+$ molecular ion (C$^{35}$Cl$^+$ and C$^{37}$Cl$^+$) were found to react with those of acetonitrile (CH$_3$CN and CD$_3$CN) to first form HNCCl$^+$ and C$_2$H$_3$$^+$ primary ionic products in nearly equal amounts, before they both react again to form the CH$_3$CNH$^+$ secondary ionic product \cite{Krohn21a,Krohn24a}.

\subsubsection{Structural isomers}

RF traps have recently also featured in recent studies investigating the effect of the different connectivities of the atoms within structural isomers on their reactivity towards sympathetically cooled ions. The allene (H$_2$C$_3$H$_2$) and propyne (HC$_3$H$_3$) structural isomers (Fig. \ref{figMS1}) were found to react differently with acetylene cations (C$_2$H$_2$$^+$) yielding isomer-specific product branching ratios: while the reaction with allene mainly led to single ionic product, C$_3$H$_3$$^+$, the one with propyne resulted in multiple primary ionic products: C$_3$H$_3$$^+$, C$_3$H$_4$$^+$ and C$_5$H$_5$$^+$ \cite{Schmid20a}. Such difference was ascribed to the dominance of long and short range interactions, respectively, in the reaction entrance channel \cite{Greenberg21a,Krohn24a}. The interplay between long-range interactions and short-range complex formation was also explored by changing the identity of the ionic reactant \cite{ZagorecMarks24a}. Besides the reactivity of isomeric reactants, the formation of isomeric products has also been subject to recent investigations. This is particularly relevant for laboratory astrochemistry, where the establishment of product branching ratios is just as important as the measurement of rate coefficients when supplying experimental data for astrochemical modelling \cite{Toscano24a}. Although mass-spectrometric product detection cannot distinguish between different isomeric products, energetic considerations and quantum-chemical calculations can help to exclude or indirectly identify the possible isomeric structure of reaction products. This was the case for the reaction between acetylene cations (C$_2$H$_2$$^+$) and acetonitrile (CH$_3$CN) where exothermic pathways were identified to two isomers of the C$_2$NH$_3$$^+$ (H$_2$CNCH$^+$ and H$_2$C$_2$NH$^+$) and of the C$_3$H$_4$$^+$ (CH$_2$CCH$_2$$^+$ and H$_3$C$_3$H$^+$) primary products \cite{Krohn23a}. 

The isomer-specific product branching ratio for the formation of HOC$^+$ and HCO$^+$ ions from the reaction of water (H$_2$O) with carbon cations (C$^+$) was determined by further reacting one of the isomeric product ions with nitrogen gas \cite{yang21a}. Titration of the products with $^{15}$N$_2$ led to the energetically less stable HOC$^+$ isomer forming $^{15}$N$_2$H$^+$ ions which can be detected independently of the unreactive HCO$^+$ given their different mass-to-charge ratio. The exothermicity of the reaction was found to induce extensive isomerization of the products, an effect that could not be determined in previous flow and beam-based measurements, since these probed the nascent product distribution. The ability to store the product ions, allowing for internal energy redistribution, radiative cooling or isomerization before detection, mimicked the collision-starved conditions in the interstellar medium and highlights the advantage of trap-based experiments for the investigation of astrochemically relevant processes. Trapping of the products also enables the study of secondary and subsequent reactions to follow complex reaction mechanisms comprising various elementary steps, as recently demonstrated in the study of a chain reaction previously assumed to form benzene, and eventually polycyclic aromatic compounds, in space \cite{Kocheril25a}.

\subsubsection{Stereoisomers}

\begin{figure}[!h]
\includegraphics[width=1\linewidth]{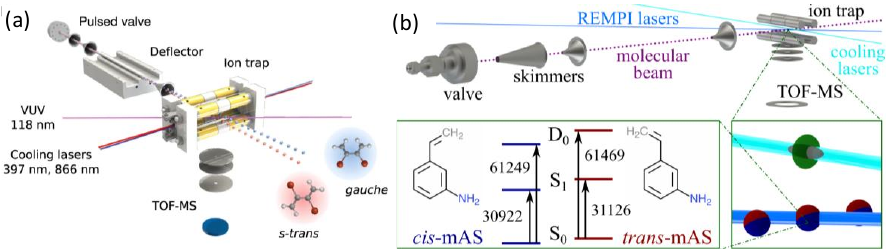}
\caption{Preparation of specific molecular configurations for isomer-specific ion--molecule reaction studies: (a) Electrostatic deflection of different conformers of polar neutral molecules in a molecular beam coupled to an ion trap \cite{Kilaj21a}, (b) generation of specific conformers of ions in a trap by conformationally selective resonance-enhanced multiphoton ionization \cite{Xu24a}.}
\label{fig5}
\end{figure}

The study of the individual reactivities of different conformational isomers (also called conformers or rotamers, Fig. \ref{figMS1})) is often hindered by their fast interconversion through rotations about single bonds under ambient conditions. The combination of adiabatic cooling and the collision-free environment provided by molecular beams can prevent this interconversion by freezing the molecules in specific conformations. If the distinct molecular geometries entail sufficiently different dipole moments in the various conformers, these can be spatially separated by application of strong inhomogeneous electric fields \cite{chang15a}. The application of such electrostatic deflection techniques has enabled the direct comparison of the individual reactivities of conformers which after separation can be individually reacted with, for instance, a target of trapped ions. 
Pioneering work employing this methodology demonstrated the importance of long-range interactions in the reactivity of the conformers of 3-aminophenol towards laser-cooled calcium ions \cite{chang13a, roesch14a}. The twofold larger rate coefficient displayed by the {\it cis} conformer with respect to the {\it trans} species was attributed to the larger dipole moment of the former, leading to stronger long-range intermolecular interactions and more efficient capture by the ion. 

More recently, the reaction of 2,3-dibromobutadiene (DBB) with calcium ions has highlighted the interplay between long and short-range effects in conformer-specific reactivity \cite{Kilaj23a}. The {\it gauche} conformer of DBB was found to react slower than the {\it s-trans}, despite having a larger dipole moment and therefore a more strongly attractive long-range interaction potential. The potential of using such controlled-molecules techniques for the elucidation of the mechanisms of complex reactions was recently demonstrated in Ref. \cite{Kilaj21a} (Fig. \ref{fig5} (a)). Here, sympathetically cooled propene ions (C$_3$H$_6$$^+$) were shown to undergo a polar cycloaddition reaction with both conformers of DBB with capture-limited rates. {\it gauche}-DBB reacted nearly twice as fast as {\it s-trans}-DBB on account of the additional ion--permanent-dipole attractive term in the long-range part of the potential, besides the ion--induced-dipole term present for both conformers. The reactivity of both conformers experimentally evidenced the coexistence of two limiting reaction mechanisms that are usually adopted to characterize cycloaddition reactions, i.e., a concerted and a stepwise mechanism. While both concerted and stepwise mechanisms were conjectured to be effective for the {\it gauche} conformer of DBB, only the stepwise pathway can connect the {\it trans} species to the products. 

Conformational control of the ionic reactant was also recently demonstrated by resonantly ionising either conformer of {\it meta}-aminostyrene (mAS, Fig. \ref{figMS1} and Fig. \ref{fig5} (b)) inside an ion trap before sympathetically cooling them within a Coulomb crystal of Ca$^+$ ions \cite{Xu24a}. Subsequently, {\it cis}-mAS$^+$ was shown to react slower than {\it trans}-mAS$^+$ with neutral calcium atoms. Finally, the isomer-specific reactivity of geometric isomers is beginning to be explored in ion traps as well. Very recently, for instance, the reactivity of 1,2-dichloroethene (DCE, Fig. \ref{figMS1}) towards Coulomb-crystallized Ca$^+$ ions was shown to be faster for the {\it cis} geometrical isomer compared to the {\it trans} one, pointing towards a broader scope of isomeric effects to be characterized in the future \cite{Xu25a}.

\section{CONCLUSIONS AND OUTLOOK}
Over the past decades, remarkable progress has been accomplished in the control of atomic and molecular ions in traps and their application in studies of ion--molecule reactions. New methodologies for the precise preparation of collision energies, quantum states, and molecular structures have thus enabled investigations into ion--molecule dynamics at an unprecedented level of detail. 

However, a number of important objectives still has to be attained. While several studies have demonstrated the preparation of the reactants in well-defined quantum states, complete state-to-state information for ion--neutral reactions, as for instance has been realized in molecular-beam studies of neutral reactions \cite{pan23a}, remains to be achieved. Detecting product quantum states is inherently more complex than preparing initial states, because reaction products are typically distributed across a wide range of energetically and dynamically accessible channels. Thus, the development of sensitive methods to probe final-state distributions will be required. The superior sensitivity achieved with the recently developed leak-out and quantum-logic spectroscopies may provide a path for such advanced experiments. 

While first glimpses of collisional quantum effects in ion--neutral collisions have recently been observed \cite{feldker20a, weckesser21a}, a full exploration of their quantum character would require even lower collision energies than have been attained so far. For this purpose, RF traps will likely have to be replaced by alternative technologies which do not suffer from micromotion in confining ions, e.g. optical traps \cite{schmidt20b}. At such low energies, emerging sophisticated approaches like shielding of collisions by dressing the reactants with optical fields \cite{anderegg21a} may become an intriguing possibility to control ion--molecule chemistry. 

Moving to ever more complex reaction systems, the study of isomeric, and in particular conformational, effects in ionic reactions remains a widely unexplored area in spite of some recent pioneering studies. The recently demonstrated ability to selectively load different conformational ions into Coulomb crystals \cite{Xu24a} opens the door to the possibility of studying fully conformationally selected reactions where the conformation of both the ionic and neutral reactant can be controlled. Moreover, with the level of control achieved in ion traps, the investigation of chiral chemistry with selected enantiomers arises as an enticing perspective in this context.

On the theory side, the accurate modeling of ion--molecule reactions beyond simple capture models \cite{tsikritea22b} remains a persistent challenge because of their open-shell and multireference character. Further efforts will be required to reach improved theoretical descriptions, in particular for larger polyatomic systems.

Achieving these objectives will represent one of the key challenges to be addressed in this thriving field over the coming years.

\section*{DISCLOSURE STATEMENT}
The authors are not aware of any affiliations, memberships, funding, or financial holdings that
might be perceived as affecting the objectivity of this review. 

\section*{ACKNOWLEDGMENTS}
P.P. is thankful for the financial support from the Georg H. Endress Foundation. We acknowledge funding from the Swiss National Science Foundation, grant nr. TMAG-2\_209193 (P.P. and S.W.), as well as for an Ambizione grant nr. PZ00P2\_208818 (J.T.)

%

\bibliographystyle{ar-style3}
\bibliography{MainBibFile, ARPC}

\end{document}